\documentclass[3p,times,twocolumn]{elsarticle}

\usepackage{ecrc}

\volume{00}

\firstpage{1}

\journalname{Nuclear Physics B Proceedings Supplement}

\runauth{W.~Buchm\"uller}

\jid{nuphbp}

\jnltitlelogo{Nuclear Physics B Proceedings Supplement}

\usepackage{amssymb}

\usepackage[figuresright]{rotating}

\begin{document}

\begin{frontmatter}



\dochead{}

\title{Leptogenesis: Theory \& Neutrino Masses}


\author{W.~Buchm\"uller}
\address{Deutsches Elektronen-Synchrotron DESY, 22607 Hamburg, Germany}

\begin{abstract}
After a brief discussion of baryon and lepton number nonconservation,
we review the status of thermal leptogenesis with GUT scale neutrino
masses, as well as low scale alternatives with keV neutrinos as dark
matter and heavy neutrino masses within the reach of the LHC.  
Recent progress towards a full quantum mechnical description of
leptogenesis is described with resonant leptogenesis as an application. 
Finally, cosmological $B$-$L$ breaking after inflation is considered
as origin of the hot early universe, generating entropy, baryon
asymmetry and dark matter. 
\end{abstract}

\end{frontmatter}


\section{Baryon \& lepton number nonconservation }
\label{sec:introduction}
\footnote{Talk given at the XXV International Conference on
  Neutrino Physics and Astrophysics, June 3-9, 2012, Kyoto, Japan}The
basis of leptogenesis \cite{Fukugita:1986hr} are the `sphaleron processes', effective
nonperturbative interactions $O_{B+L}$ of all left-handed quarks and
leptons in the Standard Model \cite{'tHooft:1976up} (cf.~Fig.~1),
which change  baryon number (B) and lepton number (L) by a multiple of three,
while preserving $B$-$L$,
\begin{eqnarray} 
 O_{B+L} &=& \prod_i (q_{Li} q_{Li} q_{Li} l_{Li}) \ , \\ 
      \Delta B &=& \Delta L\ =\ 3  N_{CS} .   
\end{eqnarray}
Here $N_{CS}$, the Chern-Simons number, is an integer characterizing
the sphaleron gauge field configuration. At high temperatures, between
the critical temperature $T_{EW}$ of the electroweak phase transition
and a maximal temperature $T_{SPH}$,
\begin{equation}
T_{EW} \sim 100~{\rm GeV} < T < T_{SPH} \sim 10^{12}~{\rm GeV} \ ,
\end{equation}
these processes are believed to be in thermal equilibrium \cite{Kuzmin:1985mm}.
Although uncontroversial among theorists, it has to be stressed that this
important phenomenon has so far not been experimentally tested! It is
therefore very interesting that the corresponding phenomenon of
chirality changing processes in strong interactions might be
observable in heavy ion collisions at the LHC 
\cite{Kharzeev:2007jp,Kalaydzhyan:2011vx}.
\begin{figure}[t]
\begin{center}
	\resizebox{4.5cm}{!}{\includegraphics{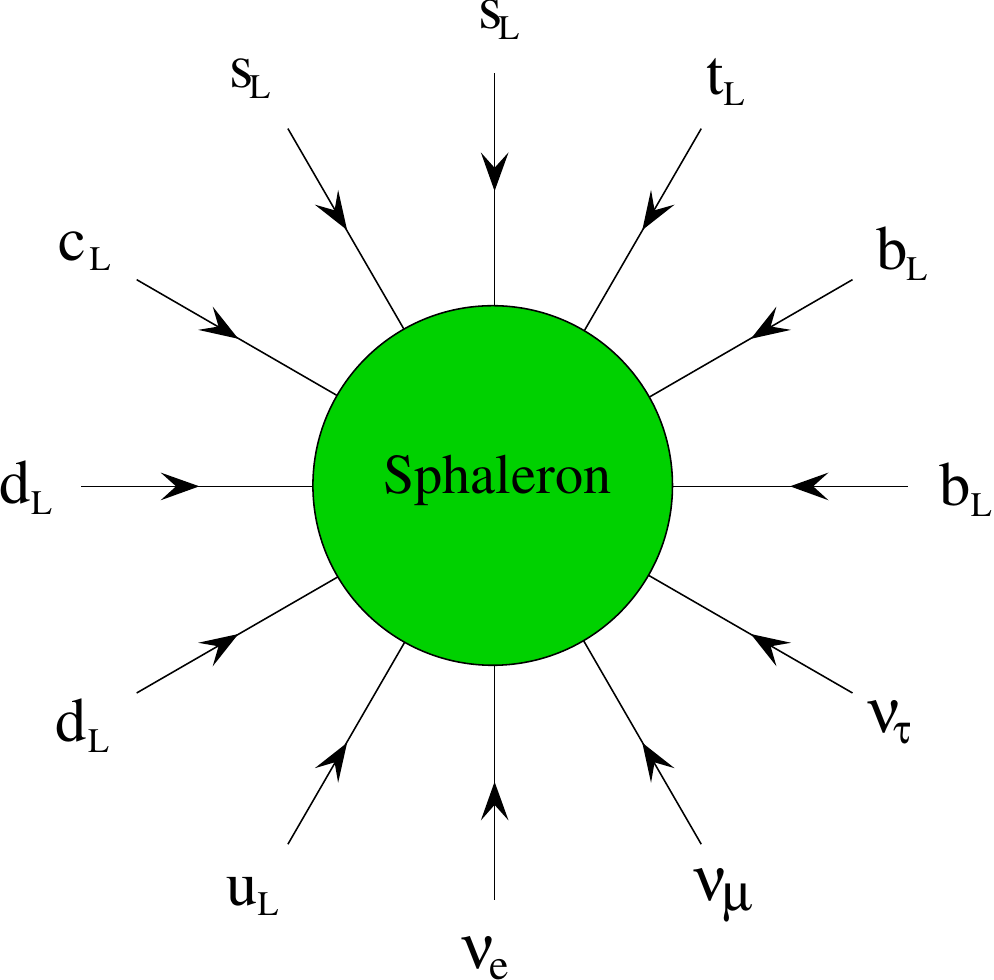}}
	\caption{One of the 12-fermion processes which are in thermal
          equilibrium in the high-temperature phase of the Standard Model.}
\end{center}
\end{figure}

Sphaleron processes relate baryon and lepton number and therefore
strongly affect the generation of the cosmological baryon asymmetry. 
Analyzing the chemical potentials of quarks and leptons in thermal
equilibrium, one obtains an important relation between the asymmetries
in $B$-,  $L$- and $B$-$L$-number,
\begin{equation}
\langle B\rangle_T = c_S \langle B-L\rangle_T 
= {c_S\over c_S-1} \langle L\rangle_T\ ,
\end{equation}
where $c_S = {\cal O}(1)$. In the Standard Model one has $c_s= 28/79$. 

This relation suggests that lepton number violation can explain
the cosmological baryon asymmetry. However, lepton number violation
 can only be weak at late
times, since otherwise any baryon asymmetry would be washed out. 
The interplay of these 
conflicting conditions leads to important contraints on neutrino properties,
and on extensions of the Standard Model in general. Because of the
sphaleron processes, lepton number violation can replace baryon number
violation in  Sakharov's conditions for baryogenesis.
\begin{figure}[t]
\begin{center}
	\resizebox{5cm}{!}{\includegraphics{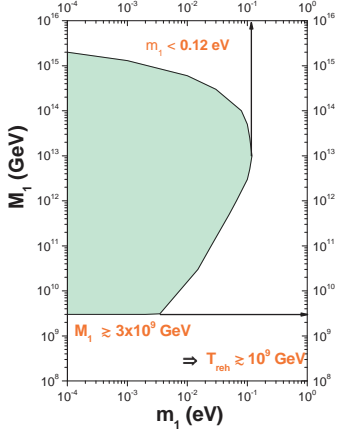}}
	\caption{Lower bounds on the smallest heavy neutrino mass $M_1$ and
          upper bounds on the smallest light neutrino mass $m_1$. From 
          Ref.~\cite{DiBari:2012fz}.}
	\label{fig:example}
\end{center}
\end{figure}

\section{Thermal leptogenesis}

Leptogenesis is an immediate consequence of the
seesaw mechanism,
which explains the smallness of light neutrino masses in terms of the
largeness of heavy Majorana neutrino masses. The heavy mass
eigenstates $N$ and the light mass eigenstates $\nu$ are given by
\begin{eqnarray}
N &\simeq& \nu_R + \nu_R^c \  : \quad m_N \simeq M \ , \\
\nu &\simeq& \nu_L + \nu_L^c \  : \quad m_\nu = - m_D{1\over M}m_D^T\ ,
\end{eqnarray}
where $m_D$ is the Dirac neutrino mass matrix. For third generation
Yukawa couplings ${\cal O}(1)$, as in some SO(10) GUT models, one
obtains the heavy and light neutrino masses,
\begin{equation}
M_3 \sim \Lambda_{\rm GUT} \sim 10^{15} {\rm GeV} , \; 
m_3 \sim {v^2\over M_3} \sim 0.01{\rm eV}\ .
\end{equation}
Remarkably, the light neutrino mass $m_3$ is compatible with 
$(\Delta m^2_{atm})^{1/2} \equiv m_{\rm atm} \simeq 0.05$~eV, as measured in
atmospheric $\nu$-oscillations. This suggests that neutrino physics probes the
mass scale of grand unification (GUT), although other interpretations of
neutrino masses are possible as well.
The heavy Majorana neutrinos have no gauge interactions. Hence, in the
early universe, they can easily be out of thermal equilibrium. This
makes $N_1$, the lightest of them, an ideal candidate for
baryogenesis, in accord with Sakharov's condition of departure from
thermal equilibrium. In the simplest form of leptogenesis the heavy
Majorana neutrinos are produced by thermal processes, which is
therefore called `thermal leptogenesis'. The $C\!P$ violating  
$N_1$ decays into lepton-Higgs pairs lead to a lepton asymmetry 
$\langle L \rangle_T \neq 0$, which is partially converted to a
baryon asymmetry $\langle B \rangle_T \neq 0$ by the sphaleron processes.
In early work on leptogenesis, it was anticipated that the light
neutrino masses are then required to have masses $m_i < \mathcal{O}(1\mathrm{eV})$ 
\cite{Buchmuller:1996pa}.
After the discovery of atmospheric neutrino oscillations, more
stringent upper bounds on neutrino masses could be derived, and
leptogenesis became increasingly popular.
\begin{figure}[t]
\begin{center}
	\resizebox{6cm}{!}{\includegraphics{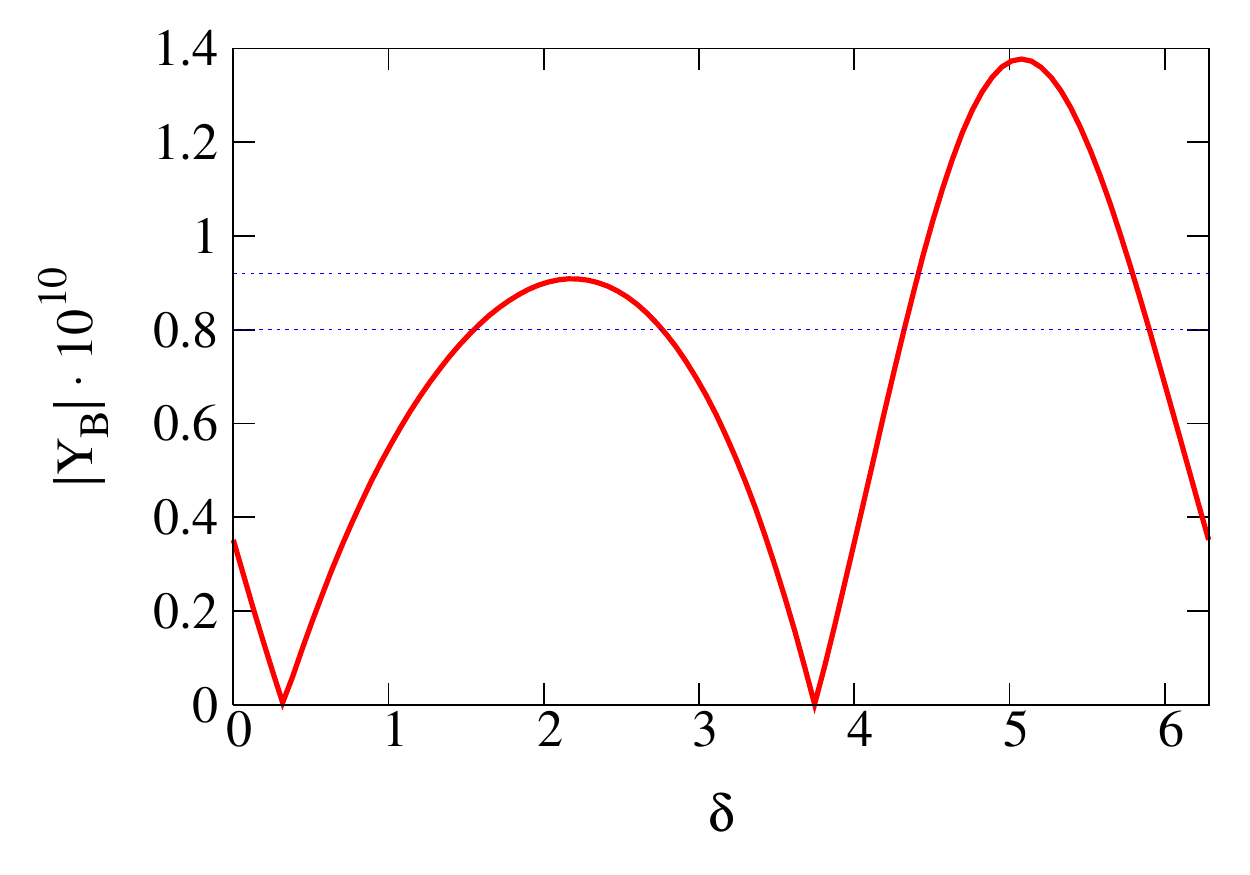}}
	\caption{Dependence of the baryon asymmetry $|Y_B|$ on the PMNS
          phase $\delta$ for a particular neutrino mass model with
          normal hierarchy. From Ref.~\cite{Molinaro:2008rg}.}
	\label{fig:example}
\end{center}
\end{figure}

The generated baryon asymmetry is proportional to the $C\!P$ 
asymmetry in $N_1$-decays. For hierarchical heavy neutrinos it
satisfies the upper bound \cite{Davidson:2002qv,Hamaguchi:2001gw} 
\begin{eqnarray}
\epsilon_1 &=& \frac{\Gamma(N_1\rightarrow l\phi)-\Gamma(N_1 \rightarrow \bar{l}\bar{\phi})}
{\Gamma(N_1 \rightarrow l\phi)+\Gamma(N_1 \rightarrow
  \bar{l}\bar{\phi})} 
\nonumber \\
& \lesssim& 10^{-6}\,{M_1\over 10^{10}\,
{\rm GeV}}\,{m_{\rm atm}\over m_1+m_3} = \epsilon_1^{\rm max}\ ,
\end{eqnarray}
which depends on the mass of $N_1$ and on $m_{\rm atm}$, the mass
splitting in atmospheric neutrino oscillations. The nonequilibrium
process of baryogenesis via leptogenesis in the hot early universe is
usually described by a set of Boltzmann equations, where also important
washout processes have to be taken into acount. Solving these
equations, one obtains lower bounds on the heavy neutrino mass $M_1$, and  
upper bounds on the light neutrino mass $m_1$ 
\cite{Buchmuller:2003gz,Nardi:2011zz}. In the simplest case of 
hierarchical heavy neutrinos, and summing over lepton flavours, one
obtains a mass window of light neutrino masses, which is favoured by
leptogenesis \cite{Buchmuller:2003gz}
\begin{equation}\label{window}
10^{-3}~{\rm eV} \lesssim m_i \lesssim 0.1~{\rm eV} \ .
\end{equation}
Note, however, that both, lower and upper bound, are modified by
lepton flavour effects, which have been extensively studied in recent
years (for reviews see, for example, Refs.~\cite{Davidson:2008bu,DiBari:2012fz}).
In view of Eq.~(\ref{window}),
knowledge of the absolute neutrino mass scale is of crucial importance.
Hence, a measurement of the neutrino masses $m_{\beta}$ in tritium
$\beta$-decay \cite{katrin} and $m_{0\nu\beta\beta}$ in
neutrinoless double $\beta$-decay \cite{0nubeta}, or the determination of 
the sum $\sum_i m_i$ from cosmology \cite{nucosmo},
consistent with Eq.~(\ref{window}), would strongly support the
leptogenesis mechanism.

How does leptogenesis depend on the phases of the Dirac and Majorana neutrino
mass matrices? Due to the large value of $\theta_{13}$ reported at
this conference \cite{dayabay}, measurement of the PMNS phase $\delta$
now appears feasible. This is certainly important, and in some models
the generated baryon asymmetry strongly depends on the phase $\delta$
(cf.~Fig.~3). In general, however, this is model
dependent\cite{Branco:2011zb}, 
and in the case of hierarchical heavy neutrinos
the effect of the PMNS phase $\delta$ is unimportant
\cite{Branco:2001pq,Davidson:2007va}.
\begin{figure}
\begin{center}
	\resizebox{7cm}{!}{\includegraphics{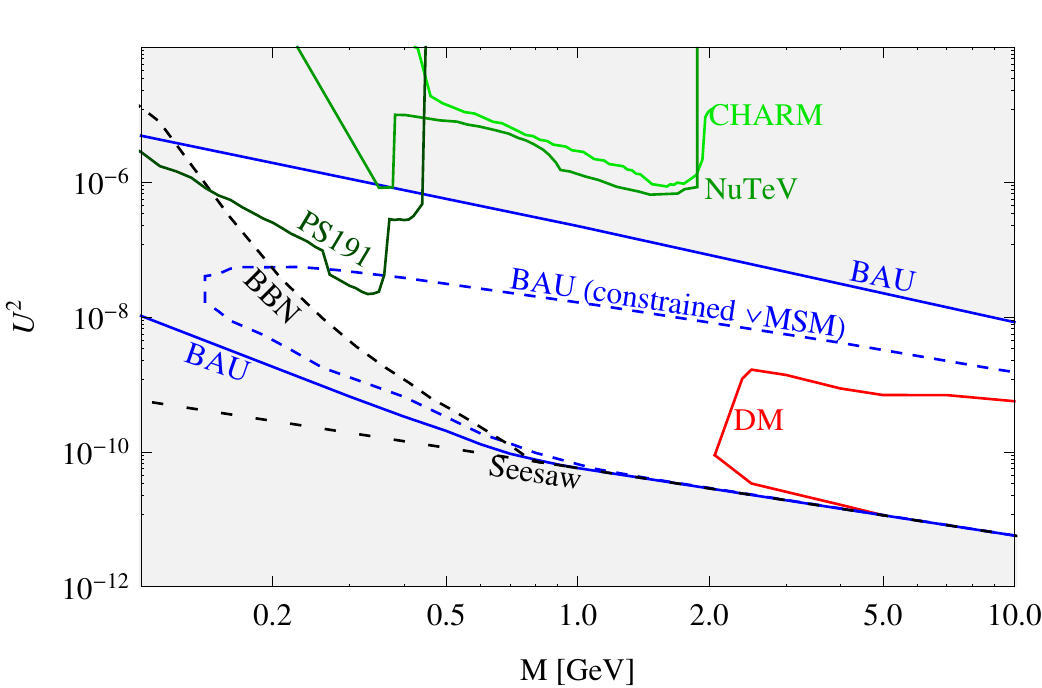}}
	\caption{Experimental and cosmological constraints on mixing
          $U^2 = {\rm tr}(\theta^\dagger\theta)$ and $N_{2,3}$ masses
          $M_{2,3} \simeq M$.  From Ref.~\cite{Canetti:2012vf}.}
	\label{fig:example}
\end{center}
\end{figure}

\section{Low Scale Alternatives}
\label{sec:Typesetting}

\subsection{'Light' Majorana Neutrinos}
\label{sec:guide}
So far we have considered right-handed neutrinos with GUT scale
Majorana masses.  In pinciple, the three right-handed (sterile)
neutrinos could have much smaller masses, of order GeV or even keV.
It is remarkable that in this case the Standard Model with three
sterile neutrinos ($\nu$MSM scenario) can account for neutrino oscillations,
baryogenesis and dark matter \cite{Asaka:2005an}.
The Standard Model Lagrangian is extended by Dirac and Majorana mass terms,
\begin{eqnarray}
\mathcal{L}_{\nu MSM} &=&\mathcal{L}_{SM} - \bar{L}_{L}F\nu_{R}\tilde{\Phi} -\bar{\nu}_{R}F^{\dagger}L_L\tilde{\Phi}^{\dagger} \nonumber\\ 
&&\quad - \frac{1}{2}(\bar{\nu_R^c}M_{M}\nu_{R} 
	+\bar{\nu}_{R}M_{M}^{\dagger}\nu^c_{R})\ ,
\end{eqnarray}
and the active-sterile mixings are described by the matrix  $\theta=m_D
M_M^{-1}$ ($U^2 = {\rm tr}(\theta^\dagger\theta)$). The scenario has
recently been studied in detail quantitatively \cite{Canetti:2012vf}. 
The lightest sterile neutrino $N_1$ provides dark matter, with a mass in
the range $1~\mathrm{keV} < M_1 \lesssim 50~\mathrm{keV}$, and tiny
mixings,  $10^{-13}\lesssim\sin^2(2\theta_{\alpha1})\lesssim 10^{-7}$, 
constrained by X-ray observations. Following Ref.~\cite{Akhmedov:1998qx},
baryogenesis is achieved by  CP-violating oscillations of $N_{2}$ and
$N_3$. To obtain the right amount of baryon asymmetry and dark matter,
resonant enhancement of CP violation is needed, with a high degeneracy
of the sterile neutrinos, $|M_2-M_3|/|M_2+M_3| \sim 10^{-11}$. The
observed dark matter abundance $\Omega_{DM}$ requires $N_{2,3}$ masses in the range
from 2-10 GeV (cf.~Fig.~4).
\begin{figure}[t]
\begin{center}
	\resizebox{7cm}{!}{\includegraphics{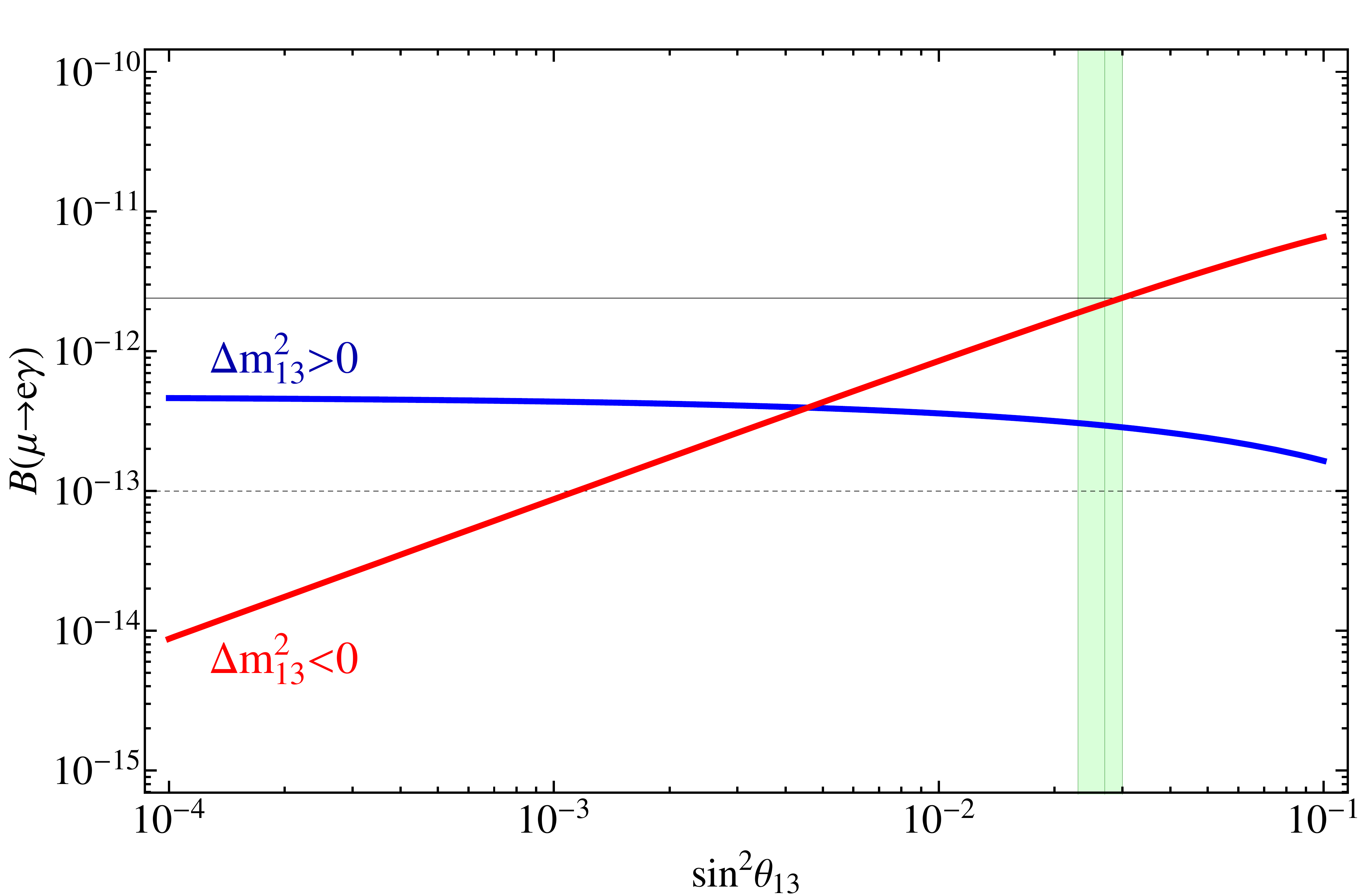}}
	\caption{Branching ratio $B(\mu\rightarrow e\gamma)$ as
          function of $\sin^2\theta_{13}$ in a scenario of resonant leptogenesis 
with neutrino masses $M_N = 120~\mathrm{GeV}$. From Ref.~\cite{Deppisch:2012zj}.}
\end{center}
\end{figure}

Sterile neutrino ($N_1$) dark matter with mass in the keV range can also be realized
in models with left-right symmetric electroweak interactions based on
the gauge group $SU(2)_L\times SU(2)_R\times U(1)$. The baryon
asymmetry is then generated by $N_2$ decays, which requires $N_{2,3}$ masses
of order $10^4 - 10^{10}$~GeV \cite{Bezrukov:2012as}.
A further 'low scale alternative' is leptogenesis at the electroweak
scale, with sterile neutrino masses $m_N \sim 100$~GeV and additional
scalar fields, which leads to specific predictions testable at the LHC
\cite{Kayser:2010fc}. More models of leptogenesis at the TeV scale can
be found in Ref.~\cite{Davidson:2008bu}.

\subsection{Resonant Leptogenesis}

As already discussed in the previous section, the seesaw mechanism
does not only work for right-handed neutrino masses at the GUT scale
and Yukawa couplings similar to quark and charged lepton Yukawa
couplings, it is also applicable for heavy neutrino masses at the TeV
scale and very small Yukawa couplings. In this case heavy neutrino
self-energy effects have to dominate the $C\!P$ asymmetry 
\cite{Flanz:1994yx,Covi:1996wh}, leading to resonant leptogenesis
in the case of quasi-degenerate right-handed neutrinos 
\cite{Pilaftsis:1997jf,Pilaftsis:2003gt}. In a particular neutrino
mass model successful leptogenesis is achieved for masses at the
electroweak scale, $M_N = 120~\mathrm{GeV}$, with a degeneracy
$\Delta M_N/ M_N \lesssim 10^{-7}$ \cite{Deppisch:2012zj}. 
It is well known that in supersymmetric models there is a close
connection between leptogenesis and lepton flavour changing processes
like $\mu \rightarrow e\gamma$ \cite{Ibarra:2009bg}.
It is interesting that in the case of resonant leptogenesis, one can
have large lepton flavour changing rates also without supersymmetry 
(cf.~Fig.~5), within the reach of the MEG experiment \cite{LFV}.

\section{Nonequilibrium theory}
\begin{figure}[b]
\begin{center}
	\resizebox{7cm}{!}{\includegraphics{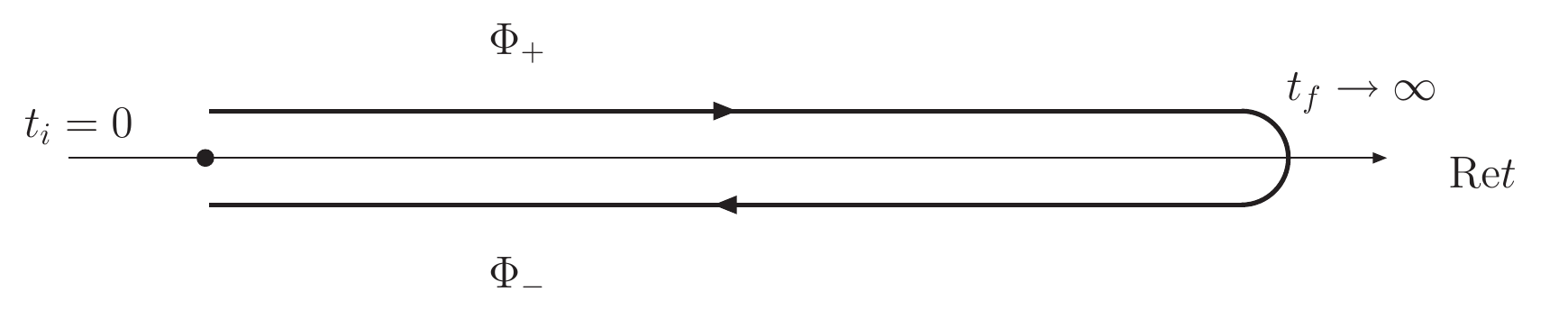}}
	\caption{Path in the complex time plane for nonequilibrium
          Green's functions.}
\end{center}
\end{figure}
Leptogenesis is a nonequilibrium process taking place in an expanding universe with
decreasing temperature. It involves quantum interferences in a crucial
manner, which implies that the standard treatment by means of Boltzmann equations
is theoretically unsatisfactory \cite{Buchmuller:2000nd}. In particular, it is currently
not possible to quote a theoretical error on the predicted bayon
asymmetry. Within quantum field theory, leptogenesis can be treated
on the basis of the Schwinger-Keldysh formalism 
\cite{Schwinger:1960qe,Keldysh:1964ud}, and during the past years
significant progress has been made towards a 'theory of leptogenesis'
\cite{Anisimov:2008dz,Garny:2009rv,Cirigliano:2009yt,Beneke:2010wd,Fidler:2011yq}.
Very important in this context is also the calculation of quantum corrections 
to decay widths and scattering cross sections at high temperature
 \cite{Besak:2010fb,Laine:2011pq,Kiessig:2011fw}.
In the Schwinger-Keldysh formalism one considers Green's functions
$\Delta$ for heavy neutrino $N_1$, lepton and Higgs on a complex time
contour starting at some initial time $t_i$ (cf.~Fig.~6).  These
Green's functions satisfy Schwinger-Dyson equations with self-energies
$\Pi_C$,
\begin{eqnarray}
\quad (\Box_1 +m^2)\Delta_C(x_{1},x_{2}) +  
\quad\qquad\qquad\quad \\ 
\int_{C}d^{4}x' \Pi_{C}(x_{1},x')
\Delta_{C}(x',x_{2}) =-i\delta_{C}(x_{1}-x_{2}) . \nonumber
\end{eqnarray}
It is then convenient to consider two particular correlation functions, the spectral
functions $\Delta^-$, which contain information about the system, and
the statistical propagators $\Delta^+$, which depends on the initial
state at time $t_i$,
\begin{eqnarray}
&&\Delta^{+}(x_{1},x_{2})
=\frac{1}{2}\langle\{\Phi(x_{1}),\Phi(x_{2})\}\rangle\ , \\
&&\Delta^{-}(x_{1},x_{2})=i\langle [\Phi(x_{1}),\Phi(x_{2})]\rangle\ .
\end{eqnarray}

\begin{figure}[t]
\begin{center}
	\resizebox{7cm}{!}{\includegraphics{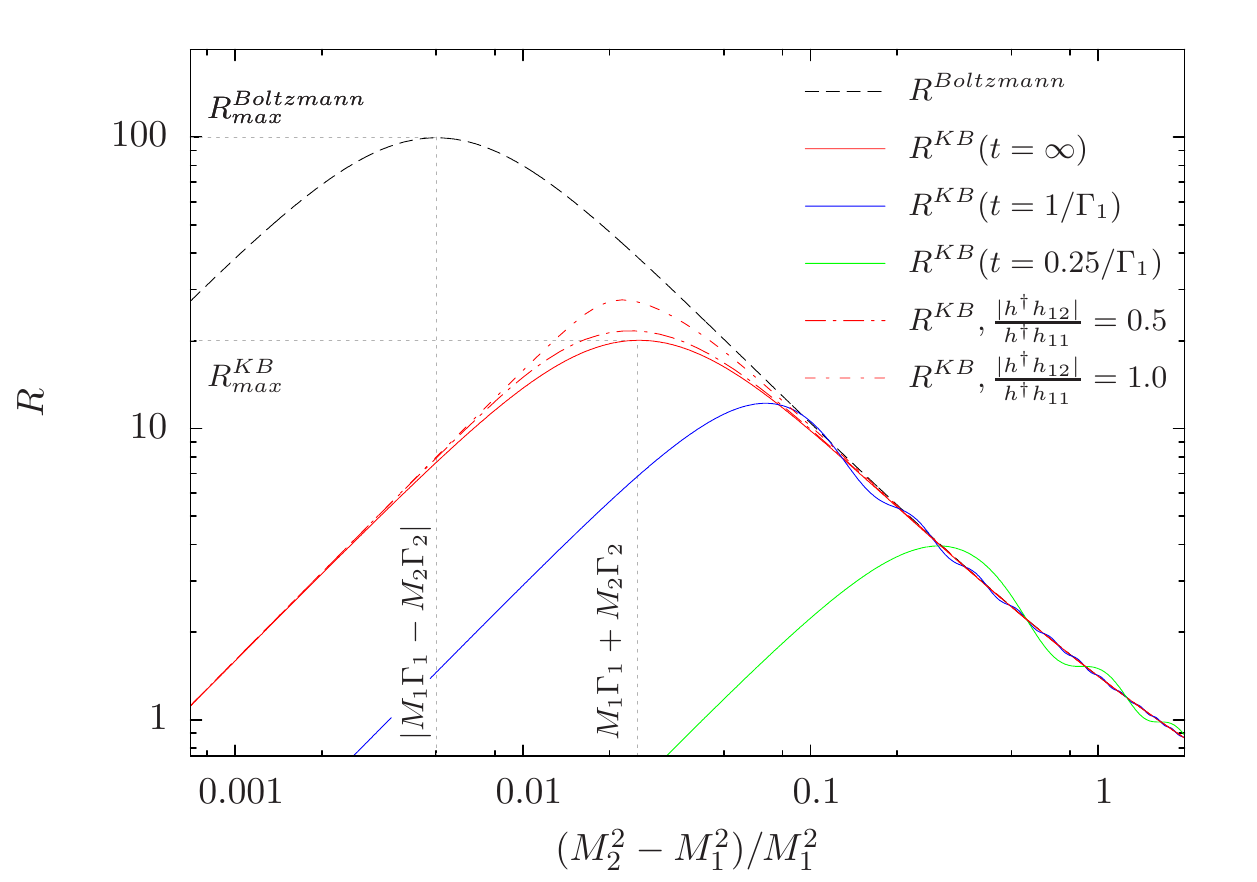}}
	\caption{Resonant enhancement $R$ as function of the
          mass-squared splitting $M_2^2-M_1^2$ for the Boltzmann
          result (dashed) and the Kadanoff-Baym result (solid lines),
          for three fixed times $t=0.25/\Gamma_1, 1/\Gamma_1,\infty$,
          and a particular choice of Yukawa couplings. From Ref.~\cite{Garny:2011hg}.}
\end{center}
\end{figure}
They satisfy the Kadanoff-Baym equations \cite{KB}
\begin{eqnarray}
\Box_{1,{\bf q}}\Delta^{-}_{\bf q}(t_{1},t_{2}) =   
- \int_{t_{2}}^{t_{1}} dt'\Pi^{-}_{\bf q}(t_{1},t')\Delta^{-}_{\bf q}(t',t_{2})\ , 
\\
\Box_{1,{\bf q}}\Delta^{+}_{\bf q}(t_{1},t_{2}) =
- \int_{t_{i}}^{t_{1}} dt'\Pi^{-}_{\bf q}(t_{1},t')\Delta^{+}_{\bf q}(t',t_{2}) \\
 +\int_{t_{i}}^{t_{2}} dt' \Pi^{+}_{\bf q}(t_{1},t')\Delta^{-}_{\bf q}(t',t_{2})\ ,
\end{eqnarray}
where we have assumed spatial homogeneity, and $\Box_{1,{\bf q}}
= (\partial^2_{t_1} + m^2 + {\bf q}^2)$ is the d'Alembert operator for a
particular momentum mode ${\bf q}$. Solving these Kadanoff-Baym
equations, one can describe the change of the system from an initial
state of zero baryon number to a final state of non-zero baryon
number, i.e. the process of baryogenesis.
Recently, this technique has been applied to resonant leptogenesis \cite{Garny:2011hg}.
The obtained effective enhancement of the $C\!P$ asymmetry is shown in
Fig.~7. The maximal enhancement predicted by Boltamann equations reads 
\begin{equation}
R^{\it Boltzmann}_{\it max} = 
\frac{M_1M_2}{2|M_1\Gamma_1  -  M_2\Gamma_2|} \ .
\end{equation}
Note that for equal masses and widths of the two heavy neutrinos $N_1$
and $N_2$, $R$ is singular, and therefore unphysical. This singularity
is cured by memory effects contained in the Kadanoff-Baym equations,
which yield the result
\begin{equation}
R^{\it KB}_{\it max} = \frac{M_1M_2}{2(M_1\Gamma_1  + M_2\Gamma_2)} \ .
\end{equation}
 In summary, the generic effect of a possible resonant enhancement of
 the $C\!P$ asymmetry is confirmed by the full quantum mechanical
 treatment. However, its size is reduced.

\section{Cosmological $B$-$L$ Breaking}

Thermal leptogenesis requires a rather large reheating temperature, 
$T_L \sim 10^{10}~\mathrm{GeV}$. In supersymmetric theories this
causes a potential problem because of the thermal production of
gravitinos, which yields the abundance
\begin{equation}
\Omega_{\tilde{G}} h^2 = C
\left(\frac{T_{\rm RH}}{10^{10}\,\textrm{GeV}}\right)
\left(\frac{100\,\textrm{GeV}}{m_{\tilde{G}}}\right)
\left(\frac{m_{\tilde{g}}}{1\,\textrm{TeV}}\right)^2 \ , 
\end{equation}
where $C \sim 0.5$,  and $T_{\rm RH}$ is the reheating temperature. For
unstable gravitinos, one has to worry about consistency with
primordial nucleosynthesis (BBN) whereas stable gravitinos may overclose the universe.
As a possible way out, nonthermal production of heavy neutrinos has
been suggested 
\cite{Lazarides:1991wu,Murayama:1992ua,Asaka:1999yd,Antusch:2010mv},
which allows to decrease the reheating temperature and therefore the
gravitino production. On the other hand, it is remarkable that for
typical gravitino and gluino masses in gravity mediated supersymmetry
breaking, a reheating temperature $T_{\rm RH} \sim 10^{10}~\mathrm{GeV}$
yields the right order of magnitude for the dark matter abundance if
the gravitino is the LSP. But why should the reheating temperature be
as large as the temperature favoured by leptogenesis, i.e. $T_{\rm RH} \sim T_L$?

It this context it is interesting to note that for typical neutrino mass parameters in
leptogenesis, $\widetilde{m}_1 \sim 0.01~\mathrm{eV}$,
$M_1 \sim 10^{10}~\mathrm{GeV}$, the heavy neutrino decay width takes
the value
\begin{equation}
\Gamma_{N_1}^0 = 
\frac{\tilde{m}_1}{8 \pi} \left(\frac{M_1}{v_\textrm{\tiny EW}}\right)^2
\sim 10^3\ \textrm{GeV} \ . 
\end{equation}
If the early universe in its evolution would reach a state where the
energy density is dominated by nonrelativistic heavy neutrinos, their
decays to lepton-Higgs pairs would lead to a relativistic plasma with temperature 
\begin{equation}
T_{\rm RH} \sim 0.2 \cdot \sqrt{\Gamma_{N_1}^0 M_P} \sim 10^{10}~\textrm{GeV} \ ,
\end{equation}
which is indeed the temperature wanted for gravitino dark matter! Is
this an intriguing hint or just a misleading coincidence?
\begin{figure}[t]
\begin{center}
	\resizebox{7cm}{!}{\includegraphics{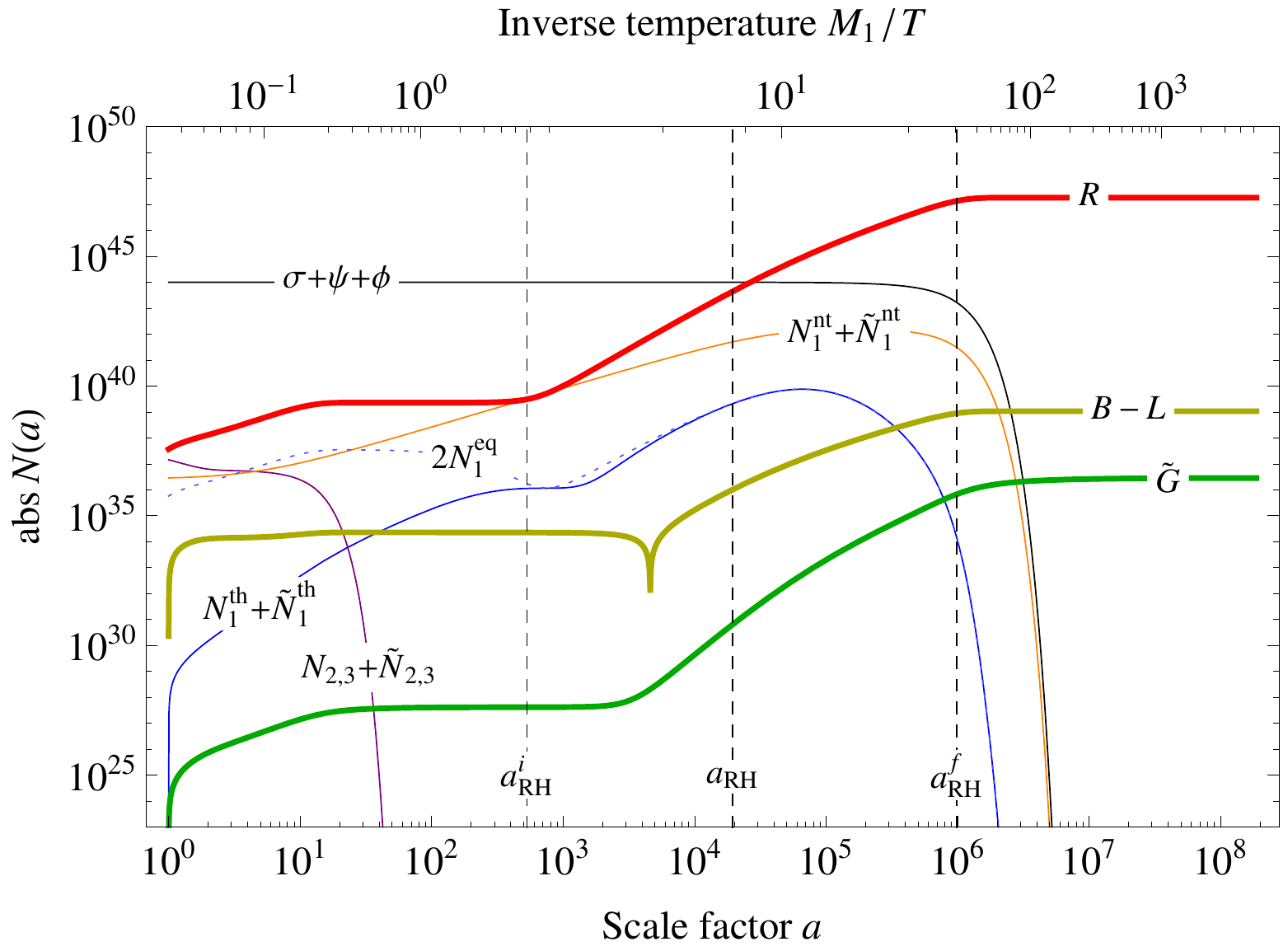}}
	\caption{Comoving number densities for particles of the
          symmetry breaking sector (Higgs $\sigma$ + higgsinos $\psi$
          + inflatons $\phi$), (non)thermally produced (s)neutrinos of
          the first generation ( $N_1^{\rm th} +\tilde{N}_1^{\rm th}$,
          $N_1^{\rm nt} +\tilde{N}_1^{\rm nt}$), (s)neutrinos of the
          second and third generaton ($N_{2,3} +\tilde{N}_{2,3}$), 
          MSSM radiation ($R$), lepton asymmetry ($B$-$L$) and
          gravitinos ($\tilde{G}$) as functions of the cosmic scale factor
          $a$. From Ref.~\cite{Buchmuller:2012wn}.}
\end{center}
\end{figure}

It is remarkable that an intermediate heavy neutrino dominance indeed
occurs in the course of the cosmological evolution if the initial
inflationary phase is driven by the false vacuum energy of unbroken
$B$-$L$ symmetry \cite{Buchmuller:2010yy,Buchmuller:2012wn}. 
Consider the supersymmetric standard model with right-handed
neutrinos, described by the superpotential (in $S\!U(5)$ notation:
 ${\bf 10} = (q,u^c,e^c)$, ${\bf 5} = (d^c,l)$),
\begin{eqnarray}
W_M = h_{ij}^u {\bf 10}_i{\bf 10}_j H_u 
          +  h_{ij}^d {\bf 5}^*_i{\bf 10}_j H_d  \nonumber \\ 
          + h_{ij}^{\nu} {\bf 5}^*_i n^c_j H_u +  h_i^n n^c_i n^c_i S_1 \ ,
\end{eqnarray}
supplemented by a term which enforces $B$-$L$ breaking,
\begin{equation}
W_{B-L} = \frac{\sqrt{\lambda}}{2} \Phi \left(v_{B-L}^2 - 2 S_1 S_2\right)\ .
\end{equation}
The Higgs fields $H_{u,d}$ and $S_{1,2}$ break electroweak symmetry
and $B$-$L$ symmetry, respectively. It is very interesting that the last term, 
$W_{B-L}$, can successfully describe inflation with $\Phi$ as inflaton
field \cite{Copeland:1994vg,Dvali:1994ms}.
Inflation
ends in tachyonic preheating where $B$-$L$ is spontaneously broken.
The false vacuum energy density is then rapidly converted into a
nonrelativistic gas of $B$-$L$ Higgs bosons and a small
admixture of a relativistic plasma of Standard Model particles
produced during preheating. Heavy
neutrinos $N_1$ are thermally produced from the plasma and nonthermally 
in decays of the $B$-$L$ Higgs bosons ($\sigma$).  This generates entropy, baryon asymmetry and
gravitino abundance.

The result of a quantitative analysis is shown in Fig.~8 for typical
parameters: $M_1 = 5.4 \times 10^{10}\,\textrm{GeV}$,
$\widetilde{m}_1 = 4.0 \times 10^{-2}\ \,\textrm{eV}$,
$m_{\widetilde{G}} = 100\ \textrm{GeV}$ and
$m_{\tilde{g}} = 1\,\textrm{TeV}$.
The final baryon asymmetry and dark matter abundance are
$\eta_B \lesssim 4 \times 10^{-9}$, 
$\Omega_{\widetilde{G}} h^2 \simeq 0.11$.
A systematic parameter scan yields a lower bound on the gravitino mass,
$10\ \textrm{GeV} \lesssim m_{\widetilde{G}}$, and a range of
heavy neutrino masses, 
$2\times 10^{10}\ \textrm{GeV} \lesssim M_1 \lesssim 2\times 10^{11}\ \textrm{GeV}$.
Note that the described mechanism for the generation of dark matter
also works for very heavy gravitinos, whose decays before BBN produce
nonthermal higgsino or wino dark matter \cite{Buchmuller:2012bt}.
\begin{figure}
\begin{center}
	\resizebox{6.5cm}{!}{\includegraphics{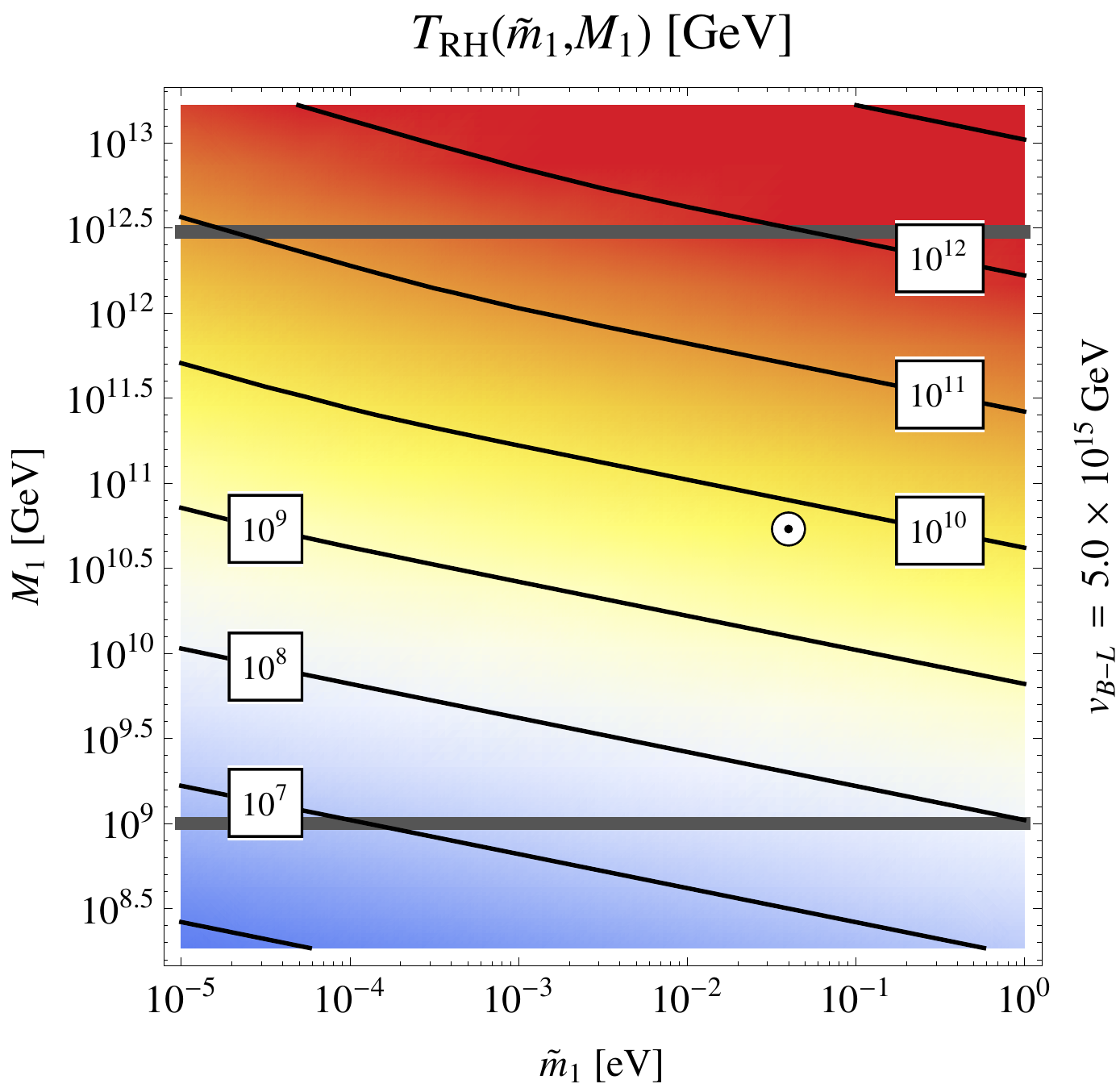}}
	\caption{Contour plot of the reheating temperature $T_{\rm RH}$
          as function of the effective light neutrino mass
          $\widetilde{m}_1$ and the heavy neutrino mass $M_1$. The
          thick gray horizontal lines represent lower and upper bound
          on $M_1$, respectively, which arise from requiring
          consistency with hybrid inflation and production of cosmic
          strings during the $B$-$L$ phase transition. The small white
          circle marks the position of the parameter point used for
          Fig.~8. From Ref.~\cite{Buchmuller:2012wn}.}
\end{center}
\end{figure}

As Fig.~8 shows, most radiation, $B$-$L$ asymmetry and gravitino
abundance are generated during a reheating period  where the cosmic scale
factor increases from $a^{i}_{\rm RH} \sim 10^3$ to $a^{f}_{\rm RH} \sim
10^6$, and the equation of state changes from matter dominance to
radiation  dominance. On the other hand, the `reheating temperature'
$T_{\rm RH}$ is roughly constant, since there is a balance between
temperature decrease due to expansion and temperature increase due 
to $B$-$L$ Higgs boson decays. Note that contrary to conventional
reheating mechanisms, the reheating temperature $T_{\rm RH}$ is now
determined by neutrino parameters! For neutrino masses consistent
with leptogenesis and dark matter, the reheating temperature varies
between $10^{7}\ \textrm{GeV}$ and $10^{12}\ \textrm{GeV}$ (cf.~Fig.~9).

\section{Summary and Outlook}

\begin{itemize}
\item Standard thermal leptogenesis is  an elegant
explanation of the cosmological matter-antimatter asymmetry. It is
consistent with GUT scale heavy neutrino masses, and it 
(successfully) predicts bounds on light neutrino masses and, in a model
dependent way, also restricts CP phases.
\item There exist various viable low-scale alternatives, with keV
  sterile neutrinos as dark matter or right-handed neutrinos with 
$\mathcal{O}(100~\mathrm{GeV})$ masses. One then assumes highly  
degenerate masses to obtain a resonance enhancement of $C\!P$
violation in heavy neutrino decays, which allows successful
leptogenesis. Contrary to GUT scale leptogenesis, these models
can be directly tested in astrophysical observations and at the LHC. 
\item Recently, considerable progress has been made in treating the 
nonequilibrium process of leptogenesis within quantum field theory,
based on the Schwinger-Keldysh formalism, with a first 
application to resonant leptogenesis. Furthermore, important quantum
corrections to decay rates and scattering processes have been
evaluated at finite temperature. One can expect that these efforts
will eventually lead to a quantitative 'theory of leptogenesis'.
\item  The possible connection between leptogenesis, dark matter and
inflation is a fascinating possibility. It is intriguing that in a
supersymmetric extension of the Standard Model with right-handed
neutrinos, a false vacuum with unbroken $B-L$ symmetry indeed provides
an initial state whose decay, via an intermediate state of heavy
neutrino dominance, can explain the observed entropy,
matter-antimatter asymmetry and dark matter abundance. 
\end{itemize}

\section*{Acknowledgments}
The author thanks Pasquale Di Bari, Valerie Domcke, Mathias Garny and
Kai Schmitz for helpful
discussions during the preparation of this talk. This work has been supported by the German Science Foundation (DFG) within 
the Collaborative Research Center 676 ``Particles, Strings and the Early
Universe''.








\end{document}